\documentclass[aps,prl,twocolumn,groupedaddress,amsmath,amssymb,groupedaddress,superscriptaddress,preprintnumbers]{revtex4-1}
\pdfoutput=1
\usepackage{graphicx}  
\usepackage{dcolumn}   
\usepackage{bm}        
\usepackage{verbatim}   
\usepackage{url}
\usepackage{hyperref}
\usepackage{multirow}
\usepackage{color}
\usepackage{textcomp}

\newcommand{\eq}[1]{eq.~\eqref{eq:#1}}

\newcommand{\fig}[1]{figure~\ref{fig:#1}}

\newcommand{\mycite}[1]{ref.~\cite{#1}}
\newcommand{\mycites}[1]{refs.~\cite{#1}}

\newcommand{\as}{\alpha_s}
\newcommand{\asvtwopi}{\Big(\frac{\alpha_s}{2\pi}\Big)}
\newcommand{\mt}{m_t}
\newcommand{\gamt}{\Gamma_t}
\newcommand{\ttb}{t\bar{t}}
\newcommand{\annlo}{NNLO${}_{\rm approx.}$}

\begin{document}

\title{Top-quark pair-production and decay at high precision}

\author{Jun Gao}
\affiliation{INPAC, Shanghai Key Laboratory for Particle Physics and Cosmology, Department
of Physics and Astronomy, Shanghai Jiao Tong University, Shanghai 200240, China}
\author{Andrew S.~Papanastasiou}
\affiliation{Cavendish Laboratory, University of Cambridge, CB3 0HE, Cambridge, UK }

\preprint{CAVENDISH-HEP-17-09}
   
\date{\today}
\begin{abstract}
We present a fully-differential and high-precision calculation of top-quark pair-production and decay at the LHC,
providing predictions for observables constructed from top-quark leptonic and $b$-flavoured jet final states. 
The calculation is implemented in a parton-level Monte Carlo and includes an approximation to the next-to-next-to leading 
order (NNLO) corrections to the production and, for the first time, the exact NNLO corrections to the decay subprocesses. 
The corrections beyond NLO are sizeable, and including them is crucial for an accurate description of the cross section 
constrained by experimental phase-space restrictions. 
We compare our predictions to published ATLAS and CMS measurements at the LHC, finding improved 
agreement compared with lower orders in the perturbative expansion. 
\end{abstract}

\maketitle

\section{Introduction}

The presence of top quarks produced in collider experiments is always inferred via the top-quark 
decay products. 
The majority of data/theory comparisons are however, performed at the level of stable tops. 
Indeed, it is through such comparisons that the top quark sector of the Standard Model is being 
carefully scrutinized. 
High statistics have been collected at all three main center-of-mass energy 
runs at the LHC and many measurements of `top-quark' inclusive cross sections and 
distributions are already down to the impressive level of a few per cent in uncertainty.
On the theoretical side, there has been equally impressive progress with top-pair production 
\cite{Baernreuther:2012ws,Czakon:2013goa,Czakon:2015owf,Czakon:2016ckf,Czakon:2016dgf,Abelof:2015lna,Czakon:2017dip,Czakon:2017wor} and 
single top \cite{Brucherseifer:2014ama,Berger:2016oht} computed fully-differentially at 
next-to-next-to leading order (NNLO) in perturbative QCD (pQCD) for stable tops. 
The fixed-order predictions can often be supplemented with various resummations which stabilize 
the predictions against the effects of large logarithms, see for example 
\mycites{Kidonakis:2000ui,Kidonakis:2001nj,Kidonakis:2003qe,Banfi:2004xa,Ahrens:2010zv,Ahrens:2011mw,
Beneke:2011mq,Cacciari:2011hy,Ferroglia:2012ku,Ferroglia:2013awa,Guzzi:2014wia,Pecjak:2016nee}.

Despite this progress, to compare with stable-top predictions, 
experiments must, in general, do two things.   
Firstly, they must extrapolate their measurements from the detector fiducial volumes out to the full 
phase space, and secondly, they must translate their measurements of final states that they are 
sensitive to, back to some definition of top-quark `partons.'
Such extrapolation and unfolding corrections are typically derived from
event generators that treat the top decay at leading order (LO), which can
lead to inconsistencies when comparing with high-precision predictions.
Moreover, the resulting systematic uncertainties can be difficult to estimate --
the observed tensions \cite{Czakon:2016olj} between the ATLAS and CMS measurements of stable-top distributions 
may just hint at some unknown systematic errors in the above procedure. 

To overcome these modelling uncertainties it is evident that experimental measurements of observables 
constructed directly from top-quark decay products in detector fiducial volumes are the quantities 
that should be compared with theoretical predictions. 
In fact, experiments have begun publishing such measurements \cite{Aad:2014kva,Khachatryan:2015mva} and 
importantly these often come with smaller systematic errors than measurements of the inclusive 
cross section \cite{Aad:2014kva}.
To fully exploit these measurements it is crucial that theoretical predictions describe the decay products 
fully differentially, and that they are as accurate as possible, i.e. 
they include higher order perturbative corrections in \emph{both} production and decay subprocesses.

Significant efforts have been made in this direction at next-to-leading order (NLO), both treating
the top-quark propagator in the narrow-width approximation (NWA) 
\cite{Bernreuther:2004jv,Melnikov:2009dn,Campbell:2012uf,Melnikov:2011ta,Melnikov:2011qx} as well as 
keeping the top quarks off their mass-shell 
\cite{Bevilacqua:2010qb,Denner:2010jp,Denner:2012yc,Falgari:2013gwa,Heinrich:2013qaa,Frederix:2013gra,
Cascioli:2013wga,Falgari:2010sf,Falgari:2011qa,Papanastasiou:2013dta,Denner:2015yca,Bevilacqua:2015qha,Bevilacqua:2016jfk,Denner:2016jyo}. 
Furthermore, frameworks have been recently developed to consistently match both sets of 
the above NLO calculations to parton showers \cite{Campbell:2014kua,Jezo:2015aia,Frederix:2016rdc,Jezo:2016ujg}. 
It is clear that these predictions come much closer to the quantities that are actually
measured by experiments.

In this letter we focus on the dominant top-quark production mode at the LHC, top-pair ($\ttb$) production,
and describe a calculation that goes beyond NLO in pQCD in both the production and 
decay stages in the NWA. 
The calculation includes an approximation of the NNLO corrections to the $\ttb$ production 
subprocess and the exact NNLO corrections to the top and antitop decays.  
We present first results of our new calculation, which has been implemented in a parton-level 
Monte Carlo, making predictions for the $\ttb$ process at the LHC in the di-lepton channel, 
fully differential in the final-state leptons, $b$-flavoured jets ($b$-jets) and missing energy.  
This represents a significant improvement to the current state-of-the-art 
at fixed order in perturbation theory and, as will be shown below, compares very favourably to 
published fiducial region measurements by the ATLAS and CMS experiments. 

\section{Details of the Calculation}

The full technical details of our calculation will be presented in a forthcoming publication, 
however we now briefly summarize the important building blocks. 
In the NWA the differential cross section for $\ttb$ production and decay in a particular decay channel 
(e.g. the di-lepton channel) can be written schematically to all orders as
\begin{align} \label{eq:nwa}
d\sigma = d\sigma_{\ttb} \times \frac{d\Gamma_{t\to b l^+ \nu_l}}{\Gamma_t} \times \frac{d\Gamma_{\bar{t}\to \bar{b} l^{'-} \bar{\nu}_{l'}}}{\Gamma_t},
\end{align}
where $d\sigma_{\ttb}$, $d\Gamma_{t\to b l^+ \nu_l}$ and $d\Gamma_{\bar{t}\to \bar{b} l^{'-} \bar{\nu}_{l'}}$ 
are the differential production cross section for a $\ttb$ pair and the differential top- and antitop-quark decay widths 
(we write the latter as $d\Gamma_t$ and $d\Gamma_{\bar{t}}$ for brevity). 
$\gamt$ is the total top-quark width. 
The `$\times$' in \eq{nwa} indicates that production and decay are combined in a way that preserves
spin-correlations. 
Each term in \eq{nwa} has a perturbative expansion in the strong coupling constant $\as$,
\begin{align} \label{eq:expansions}
d\sigma_{\ttb} &= \as^2 \sum_{i=0}^{\infty} \asvtwopi^i d\sigma^{(i)}_{\ttb}, \nonumber \\ 
d\Gamma_{t \, (\bar{t})} &= \sum_{i=0}^{\infty} \asvtwopi^i d\Gamma^{(i)}_{t \, (\bar{t})}, \;\; 
\Gamma_t = \sum_{i=0}^{\infty} \asvtwopi^i \gamt^{(i)}.
\end{align} 
In the expansion in $\as$ of \eq{nwa}, we adhere to the convention of a strict expansion, namely, we do not include
terms proportional to powers of $\as$ that are higher than the order that we work to \cite{Melnikov:2009dn,Campbell:2012uf}.   
This means that at NLO (NNLO) perturbative contributions proportional to $\as^4$ ($\as^5$) or higher are not included. 
In this convention when integrating inclusively over the decay products of the top quarks the 
cross section for the production of a stable top pair is recovered (multiplied by the appropriate branching fractions
for the $W$-boson decays).
This feature constitutes a highly non-trivial check of the implementation of each contribution
in the expansion. 

In the present calculation we include the exact NNLO corrections (i.e. corrections up to $\as^2$) in the 
expansions of $d\Gamma_{t \, (\bar{t})}$ and $\gamt$ and approximate-NNLO corrections to $d\sigma_{\ttb}$ 
(the NLO corrections $d\sigma^{(1)}_{\ttb}$ are included exactly). 
We denote our best predictions as \^NNLO (and not NNLO) since we make an approximation to  $d\sigma^{(2)}_{\ttb}$ 
(the exact NNLO corrections to the production where the spin information
of the top quarks is kept explicit, required to construct \eq{nwa}, are not currently known). 

\begin{figure}[t]
\centering
\includegraphics[trim=0.0cm 1.2cm 0.0cm 1.2cm,clip,width=0.49\textwidth]{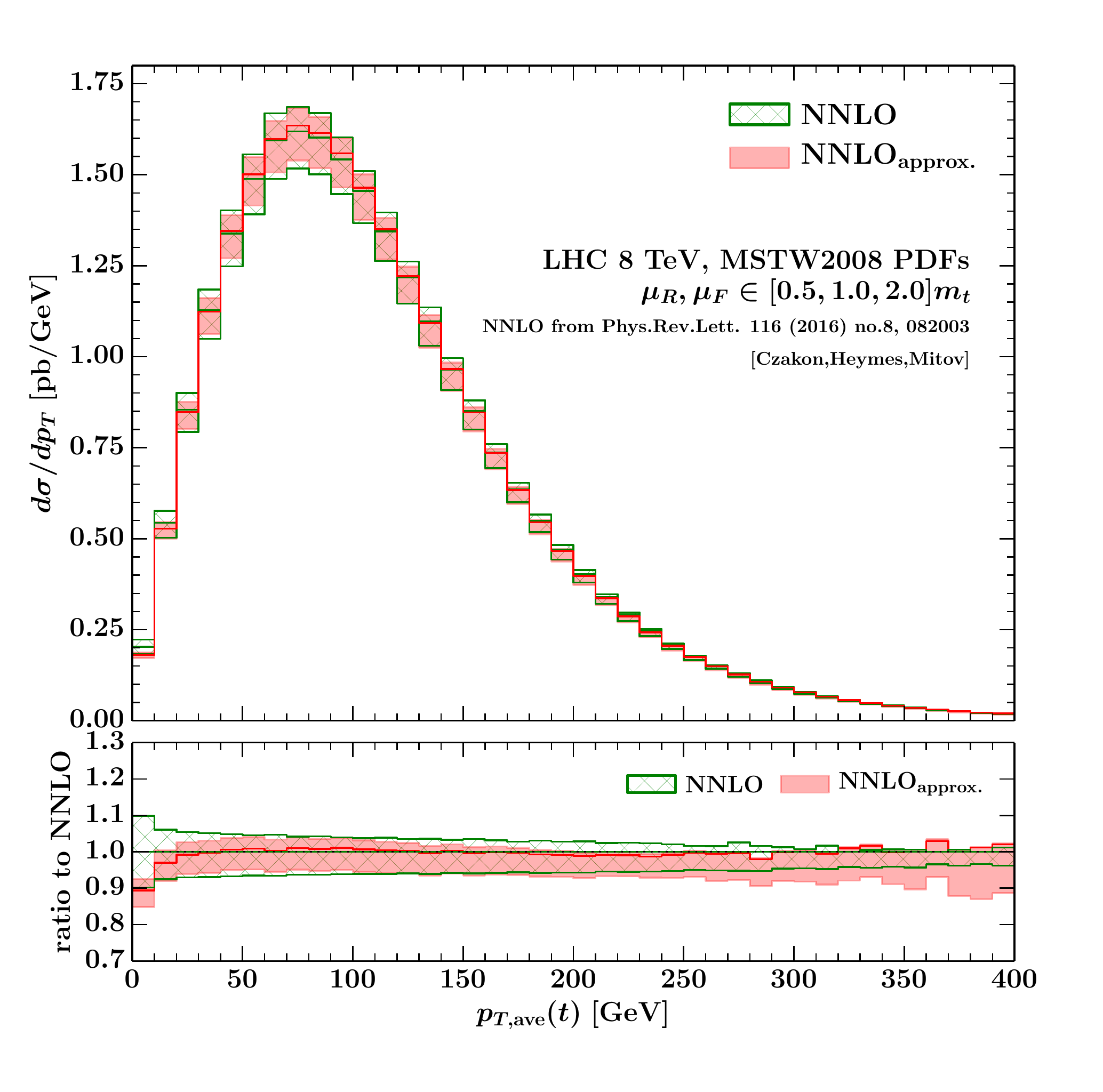} \\[-5pt]
\caption{Comparison of \annlo with exact NNLO for onshell, stable $\ttb$ production for the LHC at 8 TeV. 
Results for the exact NNLO are those published in \mycite{Czakon:2015owf}.
The NNLO uncertainty bands have been obtained via the envelope of variations of $\mu_F$ and $\mu_R$, whilst the uncertainty bands of
the \annlo~have been obtained through the envelope of scale variations \emph{and} variations of formally subleading 
contributions in $(1-z)$. See text for further details.}
\label{fig:nnlo-annlo-comparison}
\end{figure}

The approximation to the exact NNLO corrections in production builds on the work 
presented in \mycite{Broggio:2014yca}.
The starting point for this approximation is a factorization formula derived in 
the Soft-Collinear-Effective-Theory (SCET) framework \cite{Bauer:2001yt,Bauer:2000yr,Beneke:2002ph}.
It was shown in \mycite{Ahrens:2010zv} that in the soft-gluon limit 
$z = M_{\ttb}/\hat{s} = (p_t+p_{\bar t})^2/\hat{s} \to 1$ (`Pair Invariant Mass' kinematics) 
the cross section for $\ttb$ production can be written as a convolution of a hard function, 
containing the effects of virtual corrections, and a soft function, which contains
the effects of emissions of soft gluons, together with standard parton distribution functions (PDFs). 
The factorised structure makes it possible to resum large logarithms of $(1-z)$ and the subsequent 
expansion to fixed order of the resummed cross section provides an approximation to the exact NNLO.
In \mycite{Broggio:2014yca} this approach was generalised beyond the stable-top approximation and the 
spin-correlated LO decay of the top-quarks was attached to the approximate-NNLO production kernels. 
For our central prediction we take the kernels of \mycites{Broggio:2014yca} and additionally 
include terms subleading in $(1-z)$, that arise from the soft-expansion of the Altarelli-Parisi splitting functions 
\cite{Kramer:1996iq,Catani:2001ic}, which are known to bring improvements to the NNLO 
approximation \cite{Kramer:1996iq,Catani:2001ic,Broggio:2015lya,Muselli:2015kba}.
Furthermore, we explicitly use the freedom to include different subleading effects to help construct a 
reliable estimate of the theoretical uncertainty. Specifically, to make this estimate we take the envelope of 
scale variation together with variations (switching on and off) of subleading corrections of different origin
-- from the splitting functions \cite{Kramer:1996iq,Catani:2001ic} and from phase space \cite{Broggio:2015lya}.

The quality of the NNLO approximation in the production can ultimately be assessed by comparing to the exact NNLO 
cross sections for stable tops \cite{Czakon:2013goa,Czakon:2015owf}. 
We find excellent agreement with the NNLO inclusive cross section, for example, for LHC 8 TeV, $m_t=173.3$~GeV, $\mu_F=\mu_R \in [0.5,1.0,2.0]m_t$ and 
using \texttt{MSTW2008} pdfs \cite{Martin:2009iq} we have $\sigma({\rm NNLO}_{\rm approx.}) = 239.4^{+\phantom{1}5.7}_{-14.0}$~pb, 
whilst the exact NNLO cross section computed with \texttt{top++} \cite{Czakon:2011xx} is $\sigma({\rm NNLO}) = 239.2^{+\phantom{1}9.2}_{-14.8}$~pb 
(with equally good agreement for LHC 7 and 13 TeV).
Furthermore, as displayed in \fig{nnlo-annlo-comparison} for the average transverse momentum of the top and antitop,
we also find very good agreement with the exact results at the differential level 
(similar agreement is also found for the invariant mass of the top-quark pair and the average rapidity of the 
top and antitop).
These validation checks for stable top quarks provide us with confidence when using the approximate-NNLO kernels 
for the case when the top quarks are decayed. 

The NNLO corrections to the top quark decay are also calculated retaining full 
spin-correlations between production and decay.
Using the SCET-inspired phase-space slicing method presented in \mycite{Gao:2012ja},  
a small cutoff on the invariant mass ($m_j$) of all QCD partons from the top
quark decay is introduced to split the phase space in the computation of 
$d\Gamma^{(2)}_{t \, (\bar{t})}$ into resolved and unresolved regions. 
The resolved region receives contributions from the NLO corrections to the process of 
top decay plus an additional jet, and can be dealt with straightforwardly.
The contribution in the unresolved region can be factorized and calculated using SCET,
up to power corrections in $m_j^2/m_{t}^2$ \cite{Gao:2012ja}.  
The sum of resolved and unresolved contributions then converges to the exact NNLO correction 
when the cutoff is sufficiently small. 
In practice we find that a cutoff of $10^{-5}$ on $m_j^2/m_{t}^2$ is sufficient to ensure the 
remaining power corrections are negligible for all kinematic distributions considered.
We note that the NNLO decay was also computed in \mycite{Brucherseifer:2013iv}. 

Finally, the required NLO$\times$NLO production-decay ($\sim d\sigma_{\ttb}^{(1)}\times d\Gamma^{(1)}_t \times d\Gamma^{(0)}_{\bar{t}}$)
and decay-decay corrections ($\sim d\sigma_{\ttb}^{(0)}\times d\Gamma^{(1)}_t \times d\Gamma^{(1)}_{\bar{t}}$) have also been computed. 
Since production and decay subprocesses can be treated separately in the NWA, as far as their singularity structure is concerned, 
standard NLO techniques \cite{Catani:1996vz,Catani:2002hc,Frixione:1995ms} can be adapted to deal with IR-singularities in these contributions.

\section{Phenomenology}

We now apply the calculation outlined above to LHC phenomenology and in particular,
to the $\ttb$ process in the di-lepton channel.  
We first focus our attention on the fiducial cross sections measured by the ATLAS experiment
for the $e^{\pm}\mu^{\mp}$ channel at 7 and 8 TeV \cite{Aad:2014kva}, and the CMS experiment in the 
full di-lepton channel ($e^{\pm}\mu^{\mp}$, $e^+e^-$ and $\mu^+\mu^-$) at 8 TeV \cite{Khachatryan:2015mva}. 
For simplicity we compare to measurements where the indirect decays 
$W \to \tau \to e(\mu)$ are considered as backgrounds. 
The corresponding definitions of the fiducial volumes for each experiment, constructed through
cuts on final-state leptons (and $b$-jets, $J_b$), can be found in table \ref{tab:fid-xs-results}. 

The inputs for our theoretical predictions are set to
\begin{align}
\mt &= 173.3~{\rm GeV}   & \gamt^{(0)} &= 1.5048~{\rm GeV} \nonumber \\
m_W &= 80.385~{\rm GeV}  & \Gamma_W &= 2.0928~{\rm GeV} \nonumber \\
m_Z &= 91.1876~{\rm GeV} & G_F &= 1.166379\times 10^{-5}~{\rm GeV}^{-2} \nonumber
\end{align}
where we note that $\gamt^{(0)}$ is a function of $\mt$, $M_W$ and $G_F$. 
We use fixed factorization and renormalization scales~%
\footnote{In future work we intend to explore the effects of dynamical scales, which are known to improve the perturbative 
stability of the tails of distributions.}  
$\mu=\mu_F=\mu_R \in [0.5,1.0,2.0]\mt$ and vary the scale in the NLO and NNLO corrections 
to the top width $\Gamma^{(1,2)}_t(\mu)$ for consistency.
The theoretical uncertainty bands are obtained by taking the envelope of the predictions for each scale. 
For the approximate-NNLO corrections in production, as stated earlier, we additionally take the envelope of predictions
computed with different subleading terms in $(1-z)$. 
We use LO, NLO, NNLO \texttt{MMHT2014} PDFs \cite{Harland-Lang:2014zoa} with the corresponding value 
for $\alpha_s(M_Z)$ for our LO, NLO and \^NNLO predictions. 
In the results presented here we treat the $W$-bosons in the NWA. 

\begin{table*}[t]
\centering
\begin{tabular}{c@{\hskip 0.35cm} c@{\hskip 0.40cm} c@{\hskip 0.25cm} c@{\hskip 0.25cm}  c@{\hskip 0.25cm} c@{\hskip 0.40cm} c}
\hline
\hline \\[-7pt]
\multicolumn{7}{c}{ATLAS setup, $e^{\pm}\mu^{\mp}$ channel \cite{Aad:2014kva}} \\[3pt]
energy & fiducial volume & LO [pb]                     & NLO [pb]                    & \^NNLO [pb]  & $\delta_{\rm dec.}$                   & ATLAS [pb] \\[3pt]
\hline \\[-7pt]
7 TeV & $p_T(l^{\pm})>25$~GeV, $\lvert \eta(l^{\pm}) \rvert < 2.5$ & $1.592^{+39.2\%}_{-26.0\%}$ & $2.007^{+11.9\%}_{-13.2\%}$ & $2.210^{+2.2\%}_{-6.0\%}$ & -0.3\% & $2.305^{+3.8\%}_{-3.8\%}$ \\[5pt]
7 TeV & $p_T(l^{\pm})>30$~GeV, $\lvert \eta(l^{\pm}) \rvert < 2.4$ & $1.265^{+39.3\%}_{-26.1\%}$ & $1.585^{+11.8\%}_{-13.1\%}$ & $1.736^{+2.2\%}_{-6.0\%}$ & -0.8\% & $1.817^{+3.8\%}_{-3.8\%}$ \\[10pt]
8 TeV & $p_T(l^{\pm})>25$~GeV, $\lvert \eta(l^{\pm}) \rvert < 2.5$ & $2.249^{+37.9\%}_{-25.5\%}$ & $2.855^{+11.9\%}_{-12.9\%}$ & $3.130^{+2.3\%}_{-6.0\%}$ & -0.3\% & $3.036^{+4.1\%}_{-4.1\%}$ \\[5pt]
8 TeV & $p_T(l^{\pm})>30$~GeV, $\lvert \eta(l^{\pm}) \rvert < 2.4$ & $1.788^{+38.0\%}_{-25.5\%}$ & $2.256^{+11.7\%}_{-12.9\%}$ & $2.461^{+2.3\%}_{-6.1\%}$ & -0.7\% & $2.380^{+4.1\%}_{-4.1\%}$ \\[5pt]
\hline
\hline \\[-7pt]
\multicolumn{7}{c}{CMS setup, $e^{\pm}\mu^{\mp}, e^+e^-, \mu^+\mu^-$ channel \cite{Khachatryan:2015mva}, 2 $b$-jets required (anti-$k_t$ algorithm \cite{Cacciari:2008gp}, $R=0.5$)} \\[3pt]
energy & fiducial volume & LO [pb]                     & NLO [pb]                    & \^NNLO [pb]  & $\delta_{\rm dec.}$              & CMS [pb] \\[3pt]
\hline \\[-7pt]
\multirow{2}{*}{8 TeV} & $p_T(l^{\pm})>20$~GeV, $\lvert \eta(l^{\pm}) \rvert < 2.4$,  
& \multirow{2}{*}{$3.780^{+37.4\%}_{-25.3\%}$} & \multirow{2}{*}{$4.483^{+9.0\%}_{-11.5\%}$} & \multirow{2}{*}{$4.874^{+2.5\%}_{-6.8\%}$} &  \multirow{2}{*}{-8.0\%} & \multirow{2}{*}{$4.73^{+4.7\%}_{-4.7\%}$} \\[2pt] 
& $p_T(J_b)>30$~GeV, $\lvert \eta(J_b) \rvert < 2.4$ & & & & \\[3pt] 
\hline
\hline
\end{tabular}
\caption{Fiducial cross sections for a variety of LHC center-of-mass energies and setups. 
Theoretical predictions with uncertainties are tabulated at LO, NLO and \^NNLO as are the experimental measurements.
The uncertainties on the measured cross sections have been obtained by summing the individual statistical, systematic, beam
and luminosity uncertainties in quadrature.
$\delta_{\rm dec.}$ indicates the impact on the cross section of higher-order corrections to the top decay, see \eq{dec-size}.
The Monte Carlo uncertainty on all theoretical predictions is better than 1\textperthousand.}
\label{tab:fid-xs-results}
\end{table*}

Our theoretical predictions of the cross sections for the corresponding fiducial setups of ATLAS and CMS are 
tabulated in table \ref{tab:fid-xs-results}.
The ATLAS and CMS measurements are also shown in the same table. 
The errorbars on the experimental data have been obtained by summing the published individual uncertainties 
(statistical, systematic, luminosity and beam) in quadrature. 
One observes that for each setup shown, there is a reduction in the uncertainty bands with increasing perturbative
order, with the \^NNLO bands being roughly half the size of the NLO bands. 
Additionally the corrections to the cross section going from LO to NLO and from NLO to \^NNLO are also reduced, 
indicating an improved perturbative convergence. 
The corrections beyond NLO are significant, around 9-10\%, underlining  
that such corrections are crucial for an accurate description of fiducial regions~%
\footnote{We point out that large NNLO corrections in fiducial regions were also found in single top-quark production \cite{Berger:2016oht}.}.
This statement is strengthened when a comparison to the experimental measurements is made: 
for each setup considered, table \ref{tab:fid-xs-results}~reveals an improvement in the agreement 
between measurement and data. 
More precisely, use of the \^NNLO prediction brings the difference 
between the central values of theory and measurement to within 3-4\%. 

An important aspect to quantify is the size of the contributions involving corrections to the top decay. 
This can be done by studying the \% difference between the \^NNLO prediction, which includes higher-order corrections 
to the decay, and the prediction \^NNLO${}_{\rm LO \, dec.}$ that just includes the LO decay ($d\Gamma^{(0)}_{t \, (\bar{t})}$),
\begin{align} \label{eq:dec-size}
\delta_{\rm dec.} = \text{\^NNLO} / \text{\^NNLO}_{\rm LO \, dec.} - 1 \;\; [\%], 
\end{align}
also found in table \ref{tab:fid-xs-results}.
For the ATLAS setup, whose definition of the fiducial volume involves no cuts on the $b$-jets, we find 
$\delta_{\rm dec.} < 1\%$, i.e. the contributions from higher orders in the decay are very small.  
However, this is not the case for the CMS setup, where the constraint $p_T(J_b)>30$~GeV is in place,
and where the prediction that treats the top decay only at LO is 8\% larger than the 
prediction that consistently includes corrections in the decay. 
Coupled with the comparison to the precise experimental measurements, these findings point to two important conclusions.
Firstly, NNLO corrections in general are vital to describe fiducial-region cross sections accurately. 
Secondly, corrections to the production subprocess alone \emph{do not} uniformly give a good description
of the measurements -- higher-order corrections to the decay must be included to see an improved agreement 
for all setups considered. 

\begin{figure}[t]
\centering
\includegraphics[trim=0.3cm 0.0cm 0.25cm 0.7cm,clip,width=0.49\textwidth]{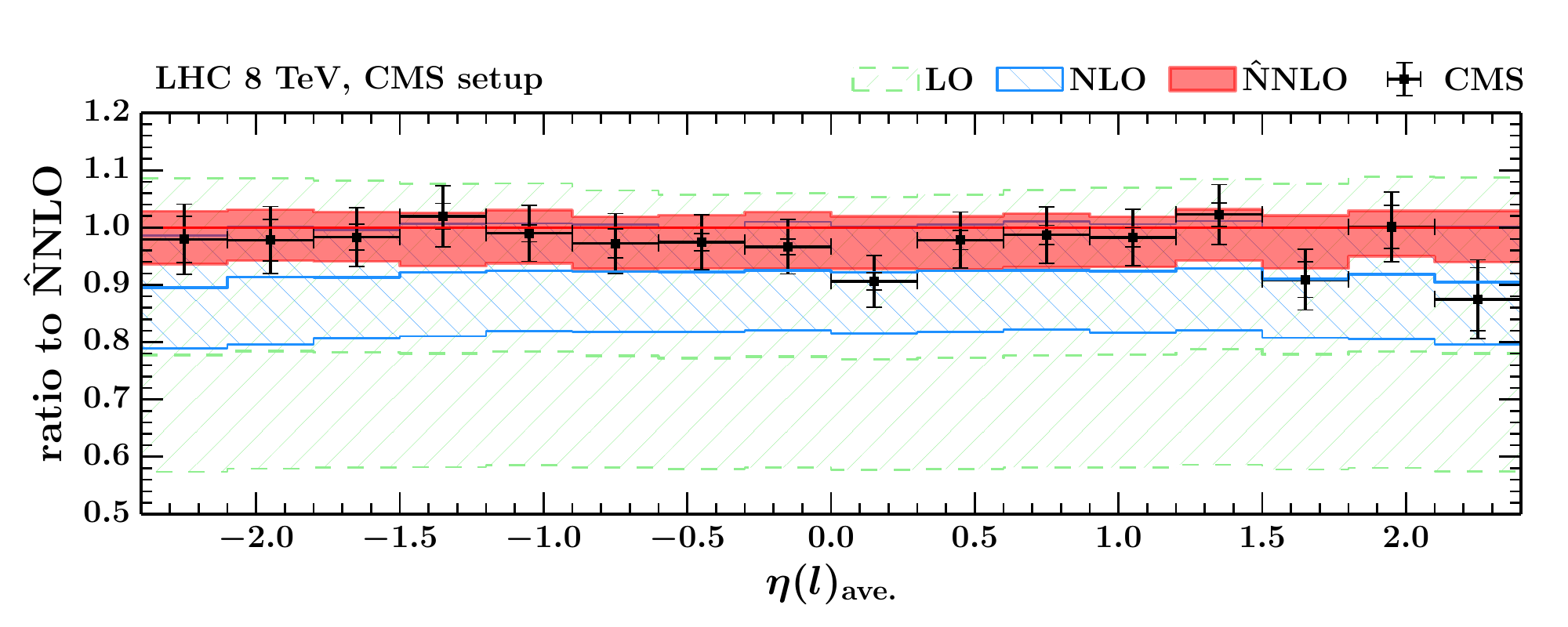} \\[-1pt]
\includegraphics[trim=0.3cm 0.0cm 0.25cm 1.3cm,clip,width=0.49\textwidth]{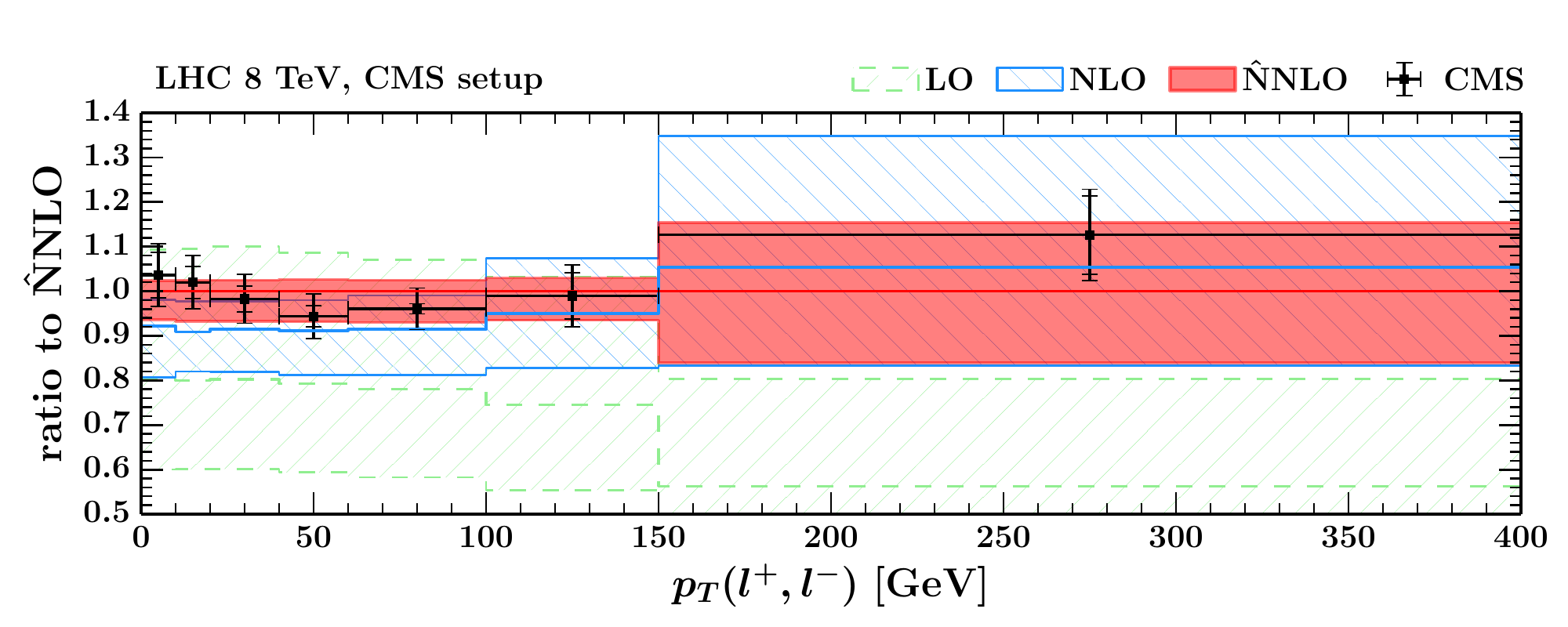} \\[-1pt]
\includegraphics[trim=0.3cm 0.0cm 0.25cm 1.3cm,clip,width=0.49\textwidth]{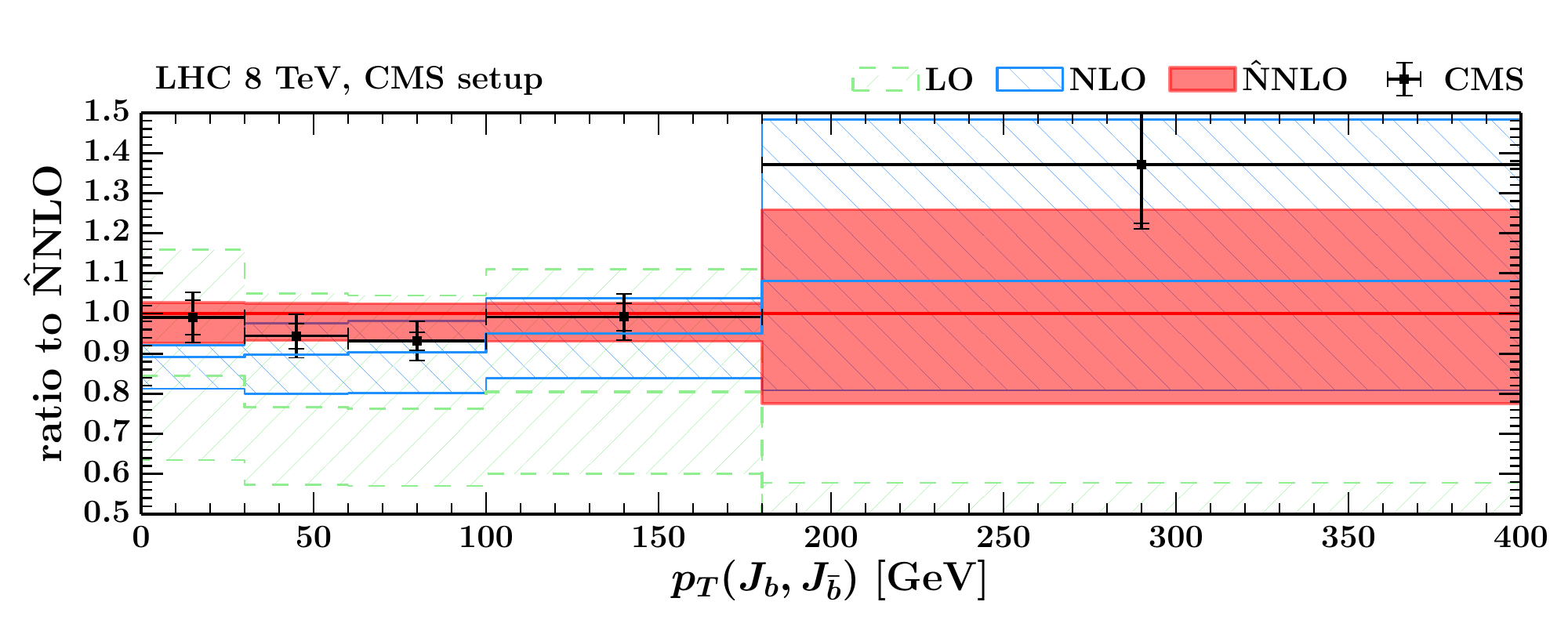} \\[-5pt]
\caption{Distributions of the average pseudo-rapidity of the leptons, $\eta(l)_{\rm ave.}$ and the 
transverse momentum of the lepton-pair, $p_T(l^+,l^-)$, and $b$-jet pair, $p_T(J_b,J_{\bar b})$. 
The plots show the CMS measurements as well as the LO, NLO and \^NNLO predictions normalized to \^NNLO. 
The errorbars and shaded bands indicate the experimental and theoretical uncertainties respectively. 
See text for further details. }
\label{fig:cms-distros}
\end{figure}

Finally, we make a comparison to differential CMS measurements \cite{Khachatryan:2015oqa} in the di-lepton channel. 
In \fig{cms-distros} we present absolute distributions for the average lepton pseudo-rapidity, $\eta(l)_{\rm ave.}$ 
and the transverse momentum of the lepton-pair, $p_T(l^+,l^-)$, and $b$-jet pair, $p_T(J_b,J_{\bar b})$, 
normalised to the \^NNLO prediction. 
We have chosen to rescale the published normalised data by the fiducial cross section found in \mycite{Khachatryan:2015mva} 
in order to make the differences between theoretical predictions at different orders more visible. 
Since there are no published uncertainties for the absolute distributions, we show the experimental points with 
two errorbars -- the smaller errorbars are those of the normalized cross section
whilst the larger ones are those of the normalized cross section added in quadrature with the uncertainty of the 
fiducial cross section used for rescaling. 
Overall, there is again good agreement between the measurements and the \^NNLO predictions -- the latter agreeing 
with the former within uncertainties in all bins.
The \^NNLO brings an improvement in the agreement not only in the overall normalization, but also in the 
shape of each distribution for the bulk of the region of phase-space measured. 
In the last bin of the $p_T(l^+,l^-)$ and $p_T(J_b,J_{\bar b})$ distributions
the agreement becomes less good, however, in these regions both theoretical and experimental uncertainty bands 
become large.  

\section{Conclusions and Outlook}

In this letter we have presented high-precision results for the fully-differential production and decay 
of a top-quark pair in fiducial regions at the LHC. 
Our results are based on the NWA and are accurate at approximate-NNLO in the production subprocess and 
exact-NNLO in the decay subprocess. 
The approximation we use in the production does an excellent job at  
approximating the exact NNLO for stable tops, giving us confidence 
in the results we present for decayed top quarks.

We have shown that, in general, the NNLO corrections are significant.
Moreover, it is vital to include corrections to the decay as well as to the production subprocess 
for an accurate description of observables constructed from top-quark decay products.
The importance of going beyond NLO is clearly seen when comparing theoretical predictions to available 
ATLAS and CMS fiducial cross section measurements. 
For different center-of-mass energies and setups we consistently find that the agreement
between theory and measurement improves when the \^NNLO predictions are used.
Additionally, we see an overall improvement in the agreement, in normalization as well as in shape
(for the bulk of the ranges considered) when comparing to distributions constructed from lepton 
and $b$-jet final states published by CMS. 

We envision that the calculation presented in this letter will open up a number of exciting 
possibilities for the study of top quarks. 
Given the impressively small experimental uncertainties on the measurements of fiducial cross sections, 
it would be particularly interesting and timely to use these measurements and exploit this new calculation
to perform an extraction of $\alpha_s$ and $\mt^{\rm pole}$.
This would bypass the need to extrapolate measurements to the full phase space and the modelling back
to top-quark partons, that affect extractions from the inclusive stable-top cross section. 
With this calculation at hand it will also be possible to quantify the impact that the exact 
NNLO top-quark decay corrections have in methods of $\mt$-extraction sensitive 
to the decay 
(see for example \mycites{Frixione:2014ala,Agashe:2016bok}). 

Data/theory comparisons of top-quark production, at the level of stable tops, have brought numerous 
impactful applications, for example, constraining the high-$x$ region of the gluon PDF \cite{Czakon:2016olj}. 
We advocate that moving towards such applications, but working with observables at the level of 
the decay products of top quarks, as we have done here, will maximize the impact that current 
and future top-quark measurements will have both within and beyond the area of top-quark physics. 

\emph{Acknowledgements}
We thank A. Broggio and A. Signer for early collaboration and many insightful discussions 
regarding the \annlo~predictions. 
We also thank E. Laenen for helpful discussions regarding subleading power terms, 
D. Heymes for clarifications regarding the exact NNLO distributions for $\ttb$
and C. Diez Pardos for making us aware of the fiducial numbers in \mycite{Khachatryan:2015mva}. 
Finally, we thank A. Signer for his comments on a draft of this letter.
The work of JG is supported by the national 1000-talent plan of China.
The work of AP is supported by the UK Science and Technology Facilities Council [grant ST/L002760/1].

\bibliographystyle{apsrev4-1}
\bibliography{tt_refs}

\end{document}